\documentclass[aps,twocolumn,nofootinbib]{revtex4}

\usepackage{graphicx}
\usepackage{dcolumn}
\usepackage{bm}

\usepackage{amssymb,amsfonts}
\usepackage{epsfig}
\usepackage{epstopdf}
\usepackage[all]{xy}
\usepackage{amsthm}
\usepackage{dcolumn}
\usepackage{hyperref}
\usepackage{url}

\newcommand{\be}{\begin{equation}}
\newcommand{\ee}{\end{equation}}
\newcommand{\bea}{\begin{eqnarray}}
\newcommand{\nn}{\nonumber}
\newcommand{\eea}{\end{eqnarray}}

\def\inbar{\,\vrule height1.5ex width.4pt depth0pt}
\def\IR{\relax{\rm I\kern-.18em R}}
\def\IC{\relax\hbox{$\inbar\kern-.3em{\rm C}$}}

\begin{document}

\title{Vacuum states for gravitons field in de Sitter space}

\author{Kazuharu Bamba\footnote{bamba@sss.fukushima-u.ac.jp}}
\affiliation{Division of Human Support System, Faculty of Symbiotic Systems Science, Fukushima University, Fukushima 960-1296, Japan}

\author{Surena Rahbardehghan\footnote{sur.rahbardehghan.yrec@iauctb.ac.ir}} \author{Hamed Pejhan\footnote{h.pejhan@piau.ac.ir}}
\affiliation{Department of Physics, Science and Research Branch, Azad University, Tehran, Iran}

\begin{abstract}
In this paper, considering the linearized Einstein equation with a two-parameter family of linear covariant gauges in de Sitter spacetime, we examine possible vacuum states for the gravitons field with respect to invariance under the de Sitter group $SO_0(1,4)$. Our calculations explicitly reveal that there exists no natural de Sitter-invariant vacuum state (the Euclidean state) for the gravitons field. Indeed, on the foundation of a rigorous group theoretical reasoning, we prove that if one insists on full covariance as well as causality for the theory, has to give up the positivity requirement of the inner product. However, one may still look for states with as much symmetry as possible, more precisely, a restrictive version of covariance by considering the gravitons field and the associated vacuum state which are, respectively, covariant and invariant with respect to some maximal subgroup of the full de Sitter group. In this regard, we treat $SO(4)$ case, and find a family of $SO(4)$-invariant states. The associated $SO(4)$-covariant quantum field is given, as well.
$$$$
\end{abstract}
\maketitle

\section{Introduction}
From the perspective of mathematical physics, de Sitter (dS) space possesses a privileged status as the unique, maximally symmetric solution to the Einstein equation with positive cosmological constant, for which, utilizing coordinates that cover the full dS manifold is needed to describe its characteristics. Moreover, the progression of observational cosmology in recent years with experiments of increasing precision like supernovae observations \cite{ARiess1009} have revealed that the Universe is in a stage of accelerated expansion. A good explanation is postulating the existence of an extra cosmic fluid, the dark energy (for reviews on the so-called dark energy and modified gravity, see, e.g., \cite{Nojiri:2010wj,Nojiri:2006ri,Book-Capozziello-Faraoni,Capozziello:2011et,delaCruzDombriz:2012xy,Bamba:2012cp,Joyce:2014kja,Koyama:2015vza,Bamba:2015uma}), for which the simplest and most convincing model is a small positive cosmological constant. Finally, examining interesting challenges which already exist at the level of quantum field theory (QFT) in de Sitter background is essential to understanding the full quantum gravity of de Sitter space.

Motivated by all of these reasons, in this work, we deal with one of the most striking aspects of de Sitter QFT which is still a source of contention in the literature, that is, the question of the existence of a state for free gravitons in dS spacetime that shares the background symmetries (In this regard, see for instance \cite{higuchi2,higuchi1,Woodard104004,Woodard1430020}). Let us be more precise. Technically, the full dS covariance of the theory implies the following requirements:
\begin{itemize}
\item{The existence of a unitary representation $\underline{U}$ of the dS group on the space of states, upon which, the field $\underline{h}(X)$ verifies
$$\underline{U}(g) \underline{h} (X) \underline{U}(g^{-1}) =  \underline{h} (gX),$$
for any $g$ in the dS group and $X$ in spacetime.}
\item{The existence of an invariant vacuum state $|0\rangle$ under the representation $\underline{U}$,
$$\underline{U}(g) |0\rangle=|0\rangle.$$}
\item{A local commutativity property, for any pair of points $X$ and $X'$ which are not causally connected,
$$ [\underline{h} (X), \underline{h} (X')]=0. $$}
\end{itemize}
On the other side, if one has to apply restricting conditions on covariance by considering the field which is covariant with regard to a subgroup of the dS group only, this is actually the so-called ``symmetry breaking".\footnote{The dS group is ten dimensional $O(1,4)$ (here, its connected component is only considered $SO_0(1,4)$), and its maximal subgroup are $O(4)$, $O(1,3)$ and $E(3)$. The first is compact and the other two are noncompact. The three subgroups correspond to transformations of dS space which leave invariant three different families of hypersurfaces. Those three families of hypersurfaces can be obtained by foliating dS space with maximally symmetric spatial surfaces. These are the standard foliations with closed ($k=1$) or open ($k=-1$) or flat ($k=0$) spatial sections \cite{SHawking1973}.} Therefore, the aforemention scientific dispute about the existence of a dS-invariant vacuum state for free gravitons field will be technically related to the covariance concept of the theory, which should be understood in terms of the action of the dS group.

On this basis, in order to examine possible vacuum states for the gravitons field, in the next section, we describe the gravitons field equation for a two-parameter family of linear covariant gauges as an eigenvalue equation of the dS group Casimir operators. The formalism is precisely introduced. In this regard, it is convenient to utilize the ambient space formalism to present the gravitons field equation in terms of the de Sitter coordinate-independent Casimir operators, which carry the group theoretical content of the theory. We briefly discuss how the occurrences of gauge invariance of the field equation leads to an indecomposable representation of the de Sitter group. In Sec. III, we define a Gupta-Bleuler triplet to manage the covariance and the gauge invariance of the theory. Thanks to the ambient space notation, an exhibition of the Gupta-Bleuler triplet for our considered field occurs in exactly the same manner as the electromagnetic field in Minkowski space. Sec. IV is devoted to constructing an invariant space of solutions under the action of $SO_0(1,4)$. The main output of our calculations is that, if we insist on full dS invariance of the theory, the positivity requirement of the inner product must be dropped. On this basis, thanks to a new representation of the canonical commutation relations, the fully dS-covariant and causal quantization of the gravitons field is presented in section V. The construction is, therefore, free of any infrared divergence. On the other side, admitting the dS symmetry breaking, we also present the set of modes of the field equation which is thoroughly $SO(4)$-invariant. The corresponding $SO(4)$-covariant quantum field then is given. Finally in section VI, we discuss our result and briefly comment on the apparent conflict of our result with the pro-invariance argument given by the mathematical physics community maintaining that there is no physical breaking of de Sitter invariance.

\section{Presentation of the de Sitter machinery}
The de Sitter spacetime is a solution of the Einstein equation with positive cosmological constant $\Lambda$. It is conveniently characterized
as a hyperboloid embedded in a 5-dimensional Minkowski spacetime
\begin{eqnarray}\label{2.1}
{M_{H}}  = \{ x \in {R}^5 ; x^2={\eta}_{\alpha\beta} {x^{\alpha}} {x^\beta} =  -H^{-2}\},
\end{eqnarray}
where $\eta_{\alpha\beta}=$ diag$(1,-1,-1,-1,-1)$ and $H$ stands for the Hubble constant.
Then, the induced metric on dS hyperboloid is as follows
\begin{eqnarray}\label{metric}
ds^2 = \eta_{\alpha\beta}dx^\alpha dx^\beta|_{x^2 = -H^{-2}} = \hat{g}_{\mu\nu}dX^\mu dX^\nu,
\end{eqnarray}
in which, the intrinsic spacetime coordinates are labeled by $X^\mu$'s and $\mu,\nu = 0, 1, 2, 3$.

This notation, namely characterizing the de Sitter spacetime as a (pseudo-)sphere in a higher-dimensional Minkowski spacetime, constitutes the ambient space approach, that contrary to a more compact intrinsic notation, makes apparent the group theoretical content of the considered model. The isometry group of the dS background is $O(1,4)$. Here, as already pointed out, only the connected component of the identity $SO_0(1,4)$ is considered.

In the ambient formalism, a tensor field ${\cal{K}}_{\alpha\beta}(x)$ can be considered as a homogeneous function in the $R^5$-variables $x^\alpha$,
\begin{equation} \label{homo}
x^\alpha \frac{\partial}{\partial x^\alpha}{\cal{K}}_{\beta\gamma}(x) = x\cdot\partial {\cal{K}}_{\beta\gamma}(x) = \sigma{\cal{K}}_{\beta\gamma}(x),
\end{equation}
in which $\sigma$ is an arbitrarily selected degree. For simplicity reasons, we consider $\sigma = 0$; the d'Alembertian operator $\square\equiv \nabla_\mu \nabla^\mu$ on de Sitter intrinsic spacetime ($\nabla_\mu$ is the covariant derivative) coincides with its counterpart $\square_5 \equiv \partial^2$ on ${R}^5$ \cite{Garidi032501}.

Due to the fact that not every homogeneous tensor field ${\cal{K}}_{\alpha\beta}(x)$ of $R^5$ represents a physical de Sitter entity, it must, in addition, verify the transversality condition to ensure that ${\cal{K}}_{\alpha\beta}(x)$ lies in the dS tangent spacetime
\begin{equation} \label{trans}
x^\alpha {\cal{K}}_{\alpha\beta}(x) = x^\beta {\cal{K}}_{\alpha\beta}(x)\;\Big(\equiv x\cdot {\cal{K}}(x)\Big) =0.
\end{equation}
The importance of this transversality property for de Sitter fields persuades us to define the symmetric, transverse projector $\theta_{\alpha\beta} = \eta_{\alpha\beta} + H^2x_\alpha x_\beta$ which allows us to construct transverse entities such as the transverse derivative,
\begin{eqnarray} \label{Partial}
\bar{\partial}_\alpha = \theta_{\alpha\beta}\partial^\beta = \partial_\alpha + H^2x_\alpha x \cdot \partial,\;\;\;x \cdot \bar{\partial} = 0.
\end{eqnarray}
In this notation, $\theta_{\alpha\beta}$ is in fact the transverse form of the dS metric in the ambient formalism,
$$\hat{g}_{\mu\nu} = \frac{\partial x^\alpha}{\partial X^\mu}\frac{\partial x^\beta}{\partial X^\nu}\theta_{\alpha\beta}.$$
Similarly, any ``intrinsic" tensor field $h_{\mu\nu}(X)$ can be locally determined by the ``transverse" tensor field ${\cal{K}}_{\alpha\beta}(x)$ as follows
\begin{equation} \label{666}
h_{\mu\nu}(X)=\frac{{\partial}x^\alpha}{{\partial}X^\mu}\frac{{\partial}x^\beta}{{\partial}X^\nu}{\cal{K}}_{\alpha\beta}(x(X)).
\end{equation}

The dS ambient space formalism allows us to express the self-adjoint $L_{\alpha\beta}$ representatives of the Killing vectors in the following form \cite{Gazeau2533,Gazeau507}
\begin{equation} \label{self-adjoint}
L_{\alpha\beta} = \Sigma_{\alpha\beta} + M_{\alpha\beta},
\end{equation}
in which $\Sigma_{\alpha\beta}$ and $M_{\alpha\beta}$ are, respectively, the action of the orbital and the spinorial parts defined as follows
\begin{eqnarray}
{\Sigma}_{\alpha\beta}{\cal{K}}_{\gamma\delta ...}\equiv-i(\eta_{\alpha\gamma}{\cal{K}}_{\beta\delta ...}\hspace{3.5cm}\nn\\
-\eta_{\beta\gamma} {\cal{K}}_{\alpha\delta ...} + \eta_{\alpha\delta}{\cal{K}}_{\gamma\beta ...}-\eta_{\beta\delta} {\cal{K}}_{\gamma\alpha ...} + ...),
\end{eqnarray}
and
\begin{equation} \label{M}
M_{\alpha\beta} \equiv -i(x_\alpha \partial_\beta - x_\beta \partial_\alpha).
\end{equation}

\begin{widetext}
Note that, admitting a system of bounded global coordinate $X^\mu$ to present a compactified version of de Sitter space ($S^3\times S$), namely
\begin{eqnarray}\label{3.2} \left \{\begin{array}{rl}
&x^0 = H^{-1}\tan \rho, \vspace{2mm}\\
&x^1 = (H\cos\rho)^{-1}(\sin\alpha\sin\theta\cos\varphi),\vspace{2mm} \\
&x^2 = (H\cos\rho)^{-1}(\sin\alpha\sin\theta\sin\varphi),\vspace{2mm}\\
&x^3 = (H\cos\rho)^{-1}(\sin\alpha\cos\theta),\vspace{2mm}\\
&x^4 = (H\cos\rho)^{-1}(\cos\alpha),
\end{array}\right.
\end{eqnarray}
$-\pi/2<\rho<\pi/2,\; 0\leq\alpha\leq\pi, \; 0\leq\theta\leq\pi$ and $0\leq\varphi<2\pi$ (the coordinate $\rho$ is timelike and acts as the conformal time), the six generators of $M_{\alpha\beta}$ associated with the compact $SO(4)$ subgroup, contracting to the Lorentz subalgebra ($H\rightarrow 0$), are given by
\begin{eqnarray}\label{M12}M_{12}=-i\frac{\partial}{\partial\varphi},\end{eqnarray}
\begin{eqnarray}\label{M32}M_{32}=-i(\sin\varphi\frac{\partial}{\partial\theta} + \cot\theta\cos\varphi\frac{\partial}{\partial\varphi}),\end{eqnarray}
\begin{eqnarray}\label{M31}M_{31}=-i(\cos\varphi\frac{\partial}{\partial\theta} + \cot\theta{\sin\varphi}\frac{\partial}{\partial\varphi}),\end{eqnarray}
\begin{eqnarray}\label{M41}M_{41}=-i(\sin\theta\cos\varphi\frac{\partial}{\partial\alpha} + \cot\alpha\cos\theta\cos\varphi\frac{\partial}{\partial\theta} - \cot\alpha\frac{\sin\varphi}{\sin\theta}\frac{\partial}{\partial\varphi}),\end{eqnarray}
\begin{eqnarray}\label{M42}M_{42}=-i(\sin\theta\sin\varphi\frac{\partial}{\partial\alpha} + \cot\alpha\cos\theta\sin\varphi\frac{\partial}{\partial\theta} + \cot\alpha\frac{\cos\varphi}{\sin\theta}\frac{\partial}{\partial\varphi}),\end{eqnarray}
\begin{eqnarray}\label{M43}M_{43}=-i(\cos\theta\frac{\partial}{\partial\alpha} - \cot\alpha{\sin\theta}\frac{\partial}{\partial\theta}),\end{eqnarray}
while considering $H\rightarrow 0$, the four generators of $M_{\alpha\beta}$ contracting to the spacetime translations are
\begin{eqnarray}\label{M01}
M_{01}=-i(\cos\rho\sin\alpha\sin\theta\cos\varphi\frac{\partial}{\partial\rho} + \sin\rho\cos\alpha\sin\theta\cos\varphi\frac{\partial}{\partial\alpha}
+ \frac{\sin\rho\cos\theta\cos\varphi}{\sin\alpha}\frac{\partial}{\partial\theta} - \frac{\sin\rho\sin\varphi}{\sin\alpha\sin\theta}\frac{\partial}{\partial\varphi}),
\end{eqnarray}
\begin{eqnarray}\label{M02}M_{02}=-i(\cos\rho\sin\alpha\sin\theta\sin\varphi\frac{\partial}{\partial\rho} + \sin\rho\cos\alpha\sin\theta\sin\varphi\frac{\partial}{\partial\alpha}
+ \frac{\sin\rho\cos\theta\sin\varphi}{\sin\alpha}\frac{\partial}{\partial\theta} + \frac{\sin\rho\cos\varphi}{\sin\alpha\sin\theta}\frac{\partial}{\partial\varphi}),
\end{eqnarray}
\begin{eqnarray}\label{M03}
M_{03}=-i(\cos\rho\sin\alpha\cos\theta\frac{\partial}{\partial\rho} + \sin\rho\cos\alpha\cos\theta\frac{\partial}{\partial\alpha}
- \frac{\sin\rho\sin\theta}{\sin\alpha}\frac{\partial}{\partial\theta}),
\end{eqnarray}
\begin{eqnarray}\label{M04}
M_{04}=-i(\cos\rho\cos\alpha\frac{\partial}{\partial\rho} - \sin\rho\sin\alpha\frac{\partial}{\partial\alpha}).
\end{eqnarray}
\end{widetext}
This categorization of generators into two sets of six and four members is essential to our investigations, and we will use it later in determining different invariant vacuum states.

Setting up the mathematical machinery, we can now proceed with the quantization of the gravitons field. We start from the Lagrangian density of pure gravity with positive cosmological constant in the dS intrinsic space,
\begin{equation} \label{L-full}
{\cal{L}}_{full}=\sqrt{-g}(R-6H^2),
\end{equation}
in which $g$ is the full metric and $R$ is the corresponding scalar curvature. By splitting the metric into a dS fixed background $\hat{g}_{\mu\nu}$ and a small fluctuation $h_{\mu\nu}$, the expanded Lagrangian to the second order in $h_{\mu\nu}$ would be
\begin{eqnarray} \label{L-expanded}
{\cal{L}}= \sqrt{-g} \Big[\frac{1}{2}\nabla_\mu h^{\mu\lambda}\nabla^\nu h_{\nu\lambda} - \frac{1}{4}\nabla_\mu h_{\nu\lambda}\nabla^\mu h^{\nu\lambda}\;\;\;\;\;\;\;\;\;\;\;\nn\\
+ \frac{1}{4}(\nabla^\mu h' - 2\nabla^\nu h^\mu_\nu)\nabla_\mu h' - \frac{1}{2}H^2 \Big(h_{\mu\nu}h^{\mu\nu} + \frac{1}{2}h'^2 \Big)\Big],
\end{eqnarray}
$h'\equiv h^\mu_\mu$. The indices are raised and lowered by $\hat{g}_{\mu\nu}$. ${\cal{L}}$ is invariant (up to a total divergence) under the gauge transformation
\begin{equation}\label{2.4} h_{\mu\nu}\rightarrow h_{\mu\nu} + (\nabla_{\mu}\Xi_{\nu} + \nabla_{\nu}\Xi_{\mu}),\end{equation}
for any vector field $\Xi_{\mu}$. As is well known, we need to break this gauge invariance for canonical quantization. Therefore, the following most general linear covariant gauge-fixing term is added to ${\cal{L}}$,
\begin{eqnarray} \label{gf}
{\cal{L}}_g = \frac{\sqrt{-g}}{2a} \Big(\nabla_\mu h^{\mu\nu} - \frac{1+b}{b}\nabla^\nu h'\Big)\;\;\;\;\;\;\;\;\;\;\;\;\; \nn\\
\times\Big(\nabla^\lambda h_{\lambda\nu} - \frac{1+b}{b}\nabla_\nu h' \Big).
\end{eqnarray}
``$a$'' and ``$b$'' are real parameters. Pursuing the least action principle, the wave equation now reads
\begin{eqnarray} \label{wave}
\Box h_{\mu\nu} + \hat{g}_{\mu\nu}(\nabla_\lambda\nabla_\rho h^{\lambda\rho} - \Box h') + \nabla_\mu\nabla_\nu h' \nn\\
- 2\nabla_{(\mu}\nabla^\lambda h_{\nu)\lambda}-H^2 (2h_{\mu\nu}+\hat{g}_{\mu\nu}h') \nn\\
+ \frac{2}{a}\Big(\nabla_{(\mu}G_{\nu)} - \frac{1+b}{b}\hat{g}_{\mu\nu}\nabla_\lambda G^\lambda \Big) = 0,
\end{eqnarray}
where $G_\nu\equiv \nabla^\lambda h_{\lambda\nu} - \frac{1+b}{b}\nabla_\nu h'$.

At this stage, using the mathematical machinery presented thus far allows us to express the field equations in terms of the coordinate-independent Casimir operators of the de Sitter group (in the Wigner sense) in analogy with the Minkowskian case \cite{PejhanII}
\begin{eqnarray}\label{2.26}
(Q_2+6){\cal{K}}+ D_2\partial_2\cdot{\cal{K}} - \frac{1}{{a}}\Big( D_2\partial_2\cdot{\cal{K}} - (\frac{1+b}{b})^2 {\cal{S}} D_1\bar\partial{\cal{K}}' \nn\\
- {(\frac{1+b}{b})} (D_2\bar\partial{\cal{K}}' - {\cal{S}}D_1\partial_2\cdot{\cal{K}}) \Big)=0,\hspace{1cm}
\end{eqnarray}
in which $Q_2$ is the second order Casimir operator of the dS group $Q_2=-\frac{1}{2}L^{\alpha\beta}L_{\alpha\beta}$,\footnote{The subscript ``$2$'' stands for the fact that the carrier space is constituted by second rank tensors.} and $\partial_2\cdot$ is the generalized divergence on the dS hyperboloid,
\begin{equation}\label{2.16} \partial_2\cdot{\cal{K}}=\partial\cdot{\cal{K}}-H^2x{\cal{K}}'-\frac{1}{2}H^2D_1{\cal{K}}', \end{equation}
$D_1=H^{-2}\bar{\partial}$, ${\cal{K}}'$ is the trace of ${\cal{K}}_{\alpha\beta}$, ${\cal{S}}$ is the symmetrizer operator (${\cal{S}}\xi_\alpha \omega_\beta=\xi_\alpha \omega_\beta+\xi_\beta \omega_\alpha$) and
\begin{equation}\label{2.15} D_2K=H^{-2}{\cal{S}}(\bar{\partial}-H^2x)K, \end{equation}
$K\equiv K_\alpha$ is an arbitrary vector field.

Note that, from now on, we only consider the traceless part of ${\cal{K}}_{\alpha\beta}$, that satisfies
\begin{eqnarray}\label{2.26TT}
(Q_2+6){\cal{K}}+ (1 - \frac{1}{{a}}) D_2\partial_2\cdot{\cal{K}} = 0.
\end{eqnarray}
It must be emphasized that, in the context of general relativity, the pure-trace sector of the graviton field does not carry any dynamics (for a detailed discussion see \cite{PejhanII}).

The above formula has a clear group-theoretical content. Indeed, utilizing the representation classification presented by the eigenvalues of the Casimir operator, one can simply associate the transverse-traceless ${\cal{K}}_{\alpha\beta}$ with a spin-2 unitary representation of the dS group. Let us make explicit this statement. The Casimir operator $Q_2$ commutes with the action of the de Sitter group generators, and therefore, it is constant in each dS unitary irreducible representation (UIR). On this basis, the eigenvalues of $Q_2$ can be considered to characterize the UIR's,
\begin{eqnarray}\label{eigen}
(Q_2 - \langle Q_2\rangle){\cal{K}}(x) = 0.
\end{eqnarray}
Respecting the notation introduced by Dixmier \cite{Dixmier}, a classification scheme utilizing a pair of parameters $(p,q)$ involved in the possible spectral values of the Casimir operators is available,
$$ Q^{(1)} = (-p(p+1) - (q+1)(q-2))I_d,$$
$$ Q^{(2)}= (-p(p+1)q(q-1))I_d.  $$
In the following, we only concentrate on the spin-2 tensor representations relevant to our present work, which are categorized as follows:
\begin{itemize}
\item{Principal series representations $(U^{2,\nu})$ (known as ``massive" representations) \cite{Gazeau304008,Flato415}
\begin{equation}\label{2.30} \langle Q_2\rangle = \nu^2 - \frac{15}{4},\;\; p=2,\;q=\frac{1}{2}+i\nu;\; \nu\in\Re.  \end{equation}}
\item{Complementary series representations $(V^{2,\mu})$
\begin{eqnarray}\label{2.31}
\langle Q_2\rangle = &&\mu - 4, \;\; p=2,\;q=\frac{1}{2}+\mu;\nn\\
&&\mu\in\Re,\;0<|\mu|<\frac{1}{2}.  \end{eqnarray}}
\item{Discrete series representations $({\Pi}^{\pm}_{2,q})$ (known as the ``massless" representations) \cite{Gazeau304008,Flato415},
\begin{eqnarray}\label{2.32}
\langle Q_2\rangle =&& - 6 - (q+1)(q-2),\nn\\
&& \;\;p=2,\; q={1},{2}.
\end{eqnarray}
Taking into account the parameter $q=1$ ($\langle Q_2\rangle =-4$) yields the representations ${\Pi}^{\pm}_{2,1}$ (in the Dixmier notation), for which, there is no Minkowskian counterpart. The second value, $q=2$ ($\langle Q_2\rangle =-6$), however, yields the representations ${\Pi}^{\pm}_{2,2}$, which are precisely the unique extensions of the massless Poincar\'{e} group representations with helicity $\pm2$.}
\end{itemize}

Here, it is worth mentioning that in obtaining the ``massless" (divergenceless and traceless) tensor field solution to the eigenvalue equation (\ref{eigen}), the calculation yields a singularity, which emerges due to the imposition of the divergencelessness condition on the tensor field \cite{Garidi3838}. This condition is required to relate the tensor field to a specific UIR of the group. The divergencelessness condition, therefore, must be dropped. Then the field equation becomes gauge invariant (see (\ref{2.26TT})) \cite{PejhanII, Garidi3838}. A direct consequence of this procedure, the appearance of the gauge invariance, is that the solution transforms under indecomposable representations of the de Sitter group. More technically, when dealing with fields involving a gauge invariance, the appearance of the Gupta-Bleuler triplet seems to be universal, and essential for quantization \cite{BinegarGUPTA,Gazeau1847}. The ambient space notation interestingly allows us to exhibit the Gupta-Bleuler triplet for the de Sitter gravitons field (the transverse-traceless sector) in precisely the same manner as it occurs for electromagnetic field \cite{PejhanI}. This triplet is based on three invariant spaces of solutions under the group action, $V_g\subset V\subset V_a$. Here, $V_a$ is the invariant space of all square integrable solutions of the field equation (according to the de Sitter-invariant indefinite inner product, that is introduced later, see (\ref{3.1})). Note that, the ``invariance", here, exactly refers to invariance under an indecomposable representation of the de Sitter group. The space $V_a$ is ``$a$'' dependent; there is a minimal value for ``$a" (=5/3)$ that removes the well-known logarithmic divergent terms \cite{PejhanII}. The physical states satisfy the divergencelessness condition and belong to an invariant subspace of the solutions, the space $V$. This invariant subspace $V$ is not invariantly complemented in $V_a$. Considering Eq. (\ref{2.26TT}), it is clearly ``$a$'' independent. The invariant subspace $V_g$ of $V$ includes the gauge solutions. It admits no invariant complement in $V$. For a more detailed discussion about the group theoretical content of the field equation (\ref{2.26TT}) (how it allows one to associate the transverse-traceless ${\cal{K}}_{\alpha\beta}$ with the dS indecomposable representation) see Refs. \cite{PejhanII,PejhanI}.

\section{Space of solutions}
The general solution to the field equation (\ref{2.26TT}) can be constructed by a combination of a scalar field and two vector fields. More exactly, considering a five-dimensional constant vector $Z_1(=Z_{1\alpha})$, a scalar field $\phi_1$ and two vector fields $K$ and $K_g$, the most general transverse-traceless, symmetric field ${\cal{K}}_{\alpha\beta}$ can be given by \cite{Gazeau2533}
\begin{eqnarray} \label{4.1}
{\cal{K}}&=&\theta\phi_1+ {\cal{S}}\bar Z_{1}K+D_{2}K_{g},\nn\\
{\cal{K}}'&=& 4\phi_1 + 2Z_{1}\cdot K + 2H^{-2}\bar\partial\cdot K_g = 0,
\end{eqnarray}
$x\cdot K= 0 = x\cdot K_g$. Applying (\ref{2.26TT}) to the above ansatz, utilizing the commutation rules and algebraic identities for the various involved operators and fields, in the case $a=5/3$ that corresponds to what we call the ``minimal case" without any logarithmic singularity,\footnote{To see the point that the value $a=5/3$ is the minimal case, one can reconsider the general case ($a\neq5/3$). The general solution then would be \cite{PejhanII}
$${\cal{K}} = {{\cal{K}}}^{(5/3)} + (5-3a) D_2 (Q_1 + 6)^{-1} (\partial\cdot {{\cal{K}}}^{(5/3)}).$$
The extra term $D_2 (Q_1 + 6)^{-1} (\partial\cdot {{\cal{K}}}^{(5/3)})$ is responsible for the appearance of a logarithmic singularity in the field solutions.} one can construct the traceless sector of the field solution ${\cal{K}}_{\alpha\beta}$ in terms of a ``massless" minimally coupled scalar field $\phi$ and a dS-invariant polarization tensor ${{\cal{D}}}_{\alpha\beta}$ (see \cite{PejhanII} for more details)
\begin{equation} \label{4.19} {{\cal{K}}}^{(5/3)}_{\alpha \beta} = {{\cal{D}}}^{(5/3)}_{\alpha \beta}(x,\partial,Z_1,Z_2)\phi, \end{equation}
where
\begin{eqnarray}\label{minimally} Q_0\phi=0, \end{eqnarray}
$Q_0=-\frac{1}{2}M_{\alpha\beta}M^{\alpha\beta}=-H^{-2}(\bar{\partial})^2$ is the scalar part of the Casimir operator and
\begin{eqnarray}
{{\cal{D}}}^{(5/3)}(x,\partial,Z_1,Z_2) = \Big( -\frac{2}{3} \theta Z_1\cdot +{\cal S}\bar Z_1 \hspace{2cm} \nn\\
+ \frac{1}{3}D_2 \Big[H^2(x\cdot Z_1) + \frac{1}{9}H^2D_1(Z_1\cdot )\Big]\Big) \nn\\
\times \Big( \bar Z_{2} - \frac{1}{2}D_1[(Z_2\cdot \bar\partial) + 2H^2(x\cdot Z_2)] \Big).\hspace{0.5cm}
\end{eqnarray}
$Z_2$ is another five-dimensional constant vector. Note that, for the sake of simplicity, from now on the index ``$5/3$'' is omitted.

Now considering Eq. (\ref{666}), one can easily convert the field solution (\ref{4.19}) into its counterpart in the bounded global coordinate $(X^\mu,\;\mu=0,1,2,3)$. In this regard, the field solution would be
\begin{eqnarray}\label{first}
h_{\mu\nu}(X)&=&\frac{\partial x^\alpha}{\partial X^\mu}\frac{\partial x^\beta}{\partial X^\nu}{\cal{K}}_{\alpha\beta}\big( x(X) \big),\nn\\
&=& \Delta_{\mu\nu}^{\lambda}(\rho,\Omega,Llm) \phi_{Llm}(\rho,\Omega)\equiv h_{\mu\nu}^{(\lambda Llm)},\;\;\;\;\;
\end{eqnarray}
in which $L=1,2,..., \;\; 0\leq l\leq L,\;\; 0\leq |m|\leq l$, the index $\lambda$ runs on all possible polarizations\footnote{There are 10 possible polarizations. One may divide them into non-zero divergent part ($\lambda=1,2,3,4$), zero divergent and zero norm part or gauge solutions ($\lambda=5,6,7,8$) and zero divergent and non-zero norm part or the central part solutions ($\lambda=9,10$). The later contains two helicities of the physical states. Here, in order to clarify the points, one may need to reconsider the Gupta-Bleuler triplet introduced in the previous section.} and
$$\Delta_{\mu\nu}^{\lambda}= \frac{\partial x^\alpha}{\partial X^\mu}\frac{\partial x^\beta}{\partial X^\nu}{\cal{D}}^{\lambda}_{\alpha\beta},$$
$\phi_{Llm}$ is the solution satisfying the equation (\ref{minimally}), that respecting the coordinate (\ref{3.2}), is given by (see Appendix)
\begin{eqnarray}\label{mini2}
\phi_{Llm}(X) = A_L(Le^{-i(L+2)\rho}+(L+2)e^{-iL\rho}) {\cal{Y}}_{Llm}(\Omega),\;\;
\end{eqnarray}
where $A_L= \frac{H}{2}[2L(L+1)(L+2)]^{-1/2}$ and the ${\cal{Y}}_{Llm}$ are the spherical harmonics on $S^3$.

De sitter spacetime is globally hyperbolic, hence, the so-called commutator $\widetilde{G}_{\mu\nu\mu'\nu'}=G^{(adv)}_{\mu\nu\mu'\nu'} - G^{(ret)}_{\mu\nu\mu'\nu'} $ is uniquely defined \cite{Isham}. Let us recall that these propagators are defined by
\begin{eqnarray}
{\cal{E}}_X^{\mu\nu\lambda\rho} G^{(adv)}_{\lambda\rho\mu'\nu'} (X,Y)= {\cal{E}}_X^{\mu\nu\lambda\rho} G^{(ret)}_{\lambda\rho\mu'\nu'}(X,Y)\hspace{1cm}\nn\\
= -\delta^{\mu\nu}_{\mu'\nu'}(X,Y),
\end{eqnarray}
where the operator ${\cal{E}}_X^{\mu\nu\lambda\rho}$ actually refers to the Euler-Lagrange field equations derived from ${\cal{L}}+{\cal{L}}_g$ (see (\ref{L-expanded}) and (\ref{gf})), so that ${\cal{E}}_X^{\mu\nu\lambda\rho}h_{\lambda\rho}=0$ leads to the wave equation (\ref{wave}), and the $\delta$-function is defined by (for any smooth symmetric tensor $f_{\mu\nu}$ on $S^4$)
$$\int\delta^{\mu\nu}_{\mu'\nu'}(X,Y)f^{\mu'\nu'}(X) d\mu(X)= f^{\mu\nu}(Y), $$
$d\mu(X) = (\cos\rho)^{-4}d\rho d\Omega$ is the $O(1,4)$-invariant measure on $M_H$. Note that, for fixed $Y$, the support in $X$ of $G^{(adv)}$ (resp. $G^{(ret)}$) lies in the past (resp. future) cone of $Y$. In this context, for (at least) any smooth solution of the field equation with compact support, we have
$$h_{\mu\nu}(\rho,\Omega)=\Big\langle(-i)\widetilde{G}_{\mu\nu\mu'\nu'}\Big((\rho,\Omega),(\rho',\Omega')\Big),h_{\mu'\nu'}(\rho',\Omega')\Big\rangle.$$
where $\langle\;,\;\rangle$ denotes the inner product which is defined for any ${h}$, ${q}$ solutions of the field equation as follows \cite{Gazeau329}
\begin{eqnarray} \label{3.1}
&\langle{h},{q}\rangle= \frac{i}{H^2}\int_{S^3,\rho=0} [({h})^*\cdot\cdot\partial_\rho{q} &\nn\\
&- 2\frac{a-1}{a}((\partial_\rho x)\cdot{({h})}^*)\cdot(\partial\cdot{q}) - (1^* \leftrightharpoons 2)]d\Omega,&\hspace{0.9cm}
\end{eqnarray}
where $a=5/3$ and ``$\cdot\cdot$" is a shortened notation for total contraction. This inner product is invariant under $\langle U_g h, U_g q \rangle= \langle{h},{q}\rangle$, in which, $U$ is the natural representation of the de Sitter group.

We shall finally use the invariant $(L^2)$ scalar product on $L^2(M_H)$ denoted by parentheses
\begin{eqnarray}
({f},{g})= \int_{M_H} f^{*}(X)\cdot\cdot g(X) d\mu(X).
\end{eqnarray}
Contrary to $\langle \;,\; \rangle$, this product is positive definite, moreover, $f$ and $g$ are not necessary solutions of the field equation.

Now, in order to construct the quantum field, one should look for a set of modes $h_{\mu\nu}^{(\lambda Llm)}$, solutions to the wave equation satisfying the following properties. Firstly,
\begin{eqnarray} \label{non dege}
&\langle h_{\mu\nu}^{(\lambda Llm)},h_{\mu\nu}^{(\lambda' L'l'm')}\rangle=\delta_{\lambda\lambda'}\delta_{LL'}\delta_{ll'}\delta_{mm'},&\nn\\
&\langle h_{\mu\nu}^{(\lambda Llm)},(h_{\mu\nu}^{(\lambda' L'l'm')})^*\rangle=0.&
\end{eqnarray}
This family can be considered to construct the ``Euclidean" vacuum in the usual terminology.

Secondly, the $h_{\mu\nu}^{(\lambda Llm)}$ and the $(h_{\mu\nu}^{(\lambda Llm)})^*$ are requested to span the space of smooth solutions to the wave equation. Given such $h_{\mu\nu}^{(\lambda Llm)}$, one considers the Hilbert space $\cal{H}_+$ they span and the corresponding bosonic Fock space $\underline{\cal{H}}_+$.\footnote{The subscript $``+"$ implies the positivity requirement of the inner product and will be clarified with more details soon.} The field $\underline{h}_{\mu\nu}(X)$ is then defined by
\begin{eqnarray}
\underline{h}_{\mu\nu}(X) = \hspace{6cm}\nn\\
\sum_{\lambda Llm} a^\lambda_{Llm} h_{\mu\nu}^{(\lambda Llm)} + \sum_{\lambda Llm}a^{\dagger\lambda}_{Llm} (h_{\mu\nu}^{(\lambda Llm)})^*,
\end{eqnarray}
where $a^\lambda_{Llm}$ and $(a^\lambda_{Llm})^\dagger$ are the usual annihilation and creation operators of the mode $h_{\mu\nu}^{(\lambda Llm)}$, respectively. Note that, this construction depends crucially on the choice made for the $h_{\mu\nu}^{(\lambda Llm)}$ or, more precisely, on the space $\cal{H}_+$ they span. To make sure this yields a physically acceptable theory, one normally requires the following additional properties of $\underline{h}_{\mu\nu}(X)$. First, $\underline{h}_{\mu\nu}(X)$ needs to be causal, actually $[\underline{h}_{\mu\nu}(X),\underline{h}_{\mu'\nu'}(X')]$ is required to equal the commutator function $-i\widetilde{G}_{\mu\nu\mu'\nu'}(X,X')$ on $M_H$ to ensure that the field satisfies the correct equal time commutation relations with its conjugate momentum. Next, one wishes all the symmetries of the classical equation to survive in the quantized theory. This means that one expects the Fock space $\underline{{\cal{H}}}_+$ to carry a unitary representation $\underline{U}$ of the isometry group of $M_H$ (and of all other symmetries of the theory),\footnote{$\underline{U}$ is actually the extension of the natural representation of the de Sitter group, $U$, to the Fock space.} and that one requires the field to transform correctly and the vacuum to be invariant. For what follows, it is of importance to recall that, in the above setting, it is sufficient to require the invariance of the solution space $\cal{H}_+$ under the natural representation of the isometry group (which extends in the obvious way to the full Fock space) to obtain the correct transformation properties of the field.

To see the point, it is convenient to define the vector valued distribution taking values in the space generated by the modes $h_{\mu\nu}^{(\lambda Llm)}$, for any real test function $f_{\mu\nu}\in D(M_H)$,\footnote{$D(M_H)$ is the space of functions $C^\infty$ with compact support in $M_H$.}
\begin{eqnarray}
X\rightarrow p_{\mu\nu}(f)(X)= \sum_{\lambda Llm} h_{\mu\nu}^{(\lambda Llm)} (X) h^{(\lambda Llm)}(f),
\end{eqnarray}
in which, $h^{(\lambda Llm)}(f)$ is the smeared form of the modes,
\begin{eqnarray}
{h}^{(\lambda Llm)}(f)& = &\int_{M_H} {h}_{\mu\nu}^{(\lambda Llm)}(X) f^{\mu\nu}(X) d\mu(X)\nn\\
&=& (h_{\mu\nu}^{(\lambda Llm)}(X),f_{\mu\nu}(X)).
\end{eqnarray}
The space generated by the $p(f)$'s is equipped with the positive invariant inner product
\begin{eqnarray}
\langle p(f) , p(g) \rangle & =& \int_{M_H} f^{\mu\nu}(X) g_{\mu\nu}(X) d\mu(X)\nn\\
&=&(f(X),g(X)).
\end{eqnarray}
Now, as usual, the field can be written as the following operator valued distribution
\begin{eqnarray}\label{pf}
\underline{h}(f) = a(p(f)) + a^\dagger(p(f)).
\end{eqnarray}
One can immediately conclude from (\ref{pf}) and the nondegeneracy condition (\ref{non dege}) that, if $\cal{H}_+$ is invariant under the action of the isometry group, then $p$ commutes with the action of de sitter group and as a result, $\underline{h}_{\mu\nu}$ also transforms correctly.

Based on the above statements, now, we are able to explain how difficulties arise in quantizing the graviton field on $M_H$. Considering the normalization constant $A_L$, one can easily see that it breaks down at $L = 0$. This is indeed the well-known ``zero-mode" problem associated with the dS massless minimally coupled scalar field \cite{AF} that leaks to the gravitons field. As a matter of fact, the space of solutions constructed by $h_{\mu\nu}^{(\lambda Llm)}(X)$ for $L\neq0$ (see (\ref{first})) does not constitute a complete set of modes. Moreover, applying the action of the dS group on these modes reveals that this set is not dS-invariant. To see the point, one can consider the following case
\begin{eqnarray}\label{L00}
(L_{03} + iL_{04}){h}_{\mu\nu}^{(\lambda, 1,0,0)}= \Big((L_{03} + iL_{04}){\Delta}^\lambda_{\mu\nu}\Big)\phi_{1,0,0}\nn\\
+ {\Delta}^\lambda_{\mu\nu}\Big((M_{03} + iM_{04})\phi_{1,0,0}\Big).
\end{eqnarray}
It is trivial that the first term for a given gauge-fixing parameter (in our case $a=5/3$) remains invariant under the group action. Reminder, it is invariant under an indecomposable representation of the dS group. However, the invariance is broken because of the last term in (\ref{L00}),
$$= ... + {\Delta}^\lambda_{\mu\nu}( -i \frac{4}{\sqrt{6}} \phi_{2,1,0} + \phi_{2,0,0} + \frac{3H}{4\pi \sqrt{6}}).$$
Obviously, the only way to prevent this symmetry breaking through the gauge-fixing procedure is the situation for which we have ${\Delta}^\lambda_{\mu\nu}=0$. It is of course the trivial solution of the field equation. Therefore, it seems that, in order to cure this symmetry breaking one should look for complementary modes to the set of solutions (\ref{first}). In this regard, we add $\Delta_{\mu\nu}^{\lambda}{\cal{C}}$ ($\cal{C}$ is a constant function) to the set of solutions and obtain the following dS-invariant space of solutions
\begin{eqnarray}\label{set0}
\{ c_0 \Delta_{\mu\nu}^{\lambda} {\cal{C}} + \sum_{\lambda Llm,\; L>0} c_{Llm} h_{\mu\nu}^{(\lambda Llm)};\; c_0, c_{Llm}\in \mathbb{C}, \nn\\
\sum_{\lambda Llm,\; L>0} |c_{Llm}|^2<\infty\}.\hspace{0.5cm}
\end{eqnarray}
Here, a crucial point should be clarified. Although, including the new mode $\Delta_{\mu\nu}^{\lambda}{\cal{C}}$ is an inevitable requirement to preserve the invariance of the theory, on the other side, however, it yields a new difficulty. More exactly, the introduced space of solutions (\ref{set0}) as an invariant inner-product space is a degenerate space,
\begin{eqnarray}
\langle\Delta_{\mu\nu}^{\lambda} {\cal{C}}, h_{\mu\nu}^{(\lambda Llm)}\rangle= \langle\Delta_{\mu\nu}^{\lambda}{\cal{C}}, \Delta_{\mu\nu}^{\lambda} {\cal{C}}\rangle=0.
\end{eqnarray}
$\Delta_{\mu\nu}^{\lambda} {\cal{C}}$ is indeed orthogonal to the whole space including itself. Therefore, due to this degeneracy, once again, canonical quantization applied to the set of modes (\ref{set0}) unavoidably leads to a noncovariant field.

To prevent this difficulty and obtain a thoroughly covariant canonical quantization of the gravitons field, by solving the equation (\ref{minimally}) directly (more exactly, (\ref{Laplace-U}) for $L=\kappa=0$), we obtain two independent solutions including the constant function mentioned above, that replace the divergent zero mode in (\ref{mini2}), as follows
\begin{eqnarray}
\phi^{(1)}_{0,0,0}= \frac{H}{2\pi} \;\;\; \mbox{and}\;\;\; \phi^{(2)}_{0,0,0}=-i\frac{H}{2\pi} (\rho + \frac{1}{2}\sin 2\rho).
\end{eqnarray}
On this basis, we present the following definition for $h_{\mu\nu}^{(\lambda,0,0,0)}$
\begin{eqnarray}\label{0-mode}
&h_{\mu\nu}^{(\lambda,0,0,0)} = \Delta_{\mu\nu}^{\lambda}(\rho,\Omega)\phi_{0,0,0},&\nn\\
&\phi_{0,0,0}= \phi^{(1)}_{0,0,0}+ {\phi^{(2)}_{0,0,0}} / {2}.&
\end{eqnarray}
Note that, the constants of normalization are chosen in order to have $\langle h_{\mu\nu}^{(\lambda,0,0,0)},h_{\mu\nu}^{(\lambda,0,0,0)}\rangle=1$. Considering this new definition, interestingly, a complete set of positive norm modes $h_{\mu\nu}^{(\lambda Llm)}$ ($L\geq 0$) is available. However, once again dS invariance breaks due to this mode. See for instance
\begin{eqnarray}\label{L11}
&(L_{03} + iL_{04}){h}_{\mu\nu}^{(\lambda,0,0,0)} = \Big((L_{03} + iL_{04}){\Delta}^{\lambda}_{\mu\nu}\Big)\phi_{0,0,0}&\nn\\
&+ {\Delta}^\lambda_{\mu\nu}\Big((M_{03} + iM_{04})\phi_{0,0,0}\Big). &
\end{eqnarray}
As mentioned before, the first term is trivially invariant under the group action. The invariance, however, is broken owing the second term,
\begin{eqnarray}\label{L11'}
&{\Delta}^\lambda_{\mu\nu}\Big((M_{03} + iM_{04})\phi_{0,0,0}\Big)={\Delta}^\lambda_{\mu\nu}\Big((M_{03} + iM_{04}){\phi^{(2)}_{0,0,0}}\Big)&\nn\\
&={\Delta}_{\mu\nu}^{\lambda} (\frac{-\sqrt{6}}{4}) \Big(i\phi_{1,0,0} +i \phi^*_{1,0,0} +\phi_{1,1,0} +\phi^*_{1,1,0}\Big)&\nn\\
&=(\frac{-\sqrt{6}}{4}) \Big(ih_{\mu\nu}^{(\lambda,1,0,0)} +i (h_{\mu\nu}^{(\lambda,1,0,0)})^* \hspace{2cm}&\nn\\
& \hspace{2cm} + h_{\mu\nu}^{(\lambda,1,1,0)} + (h_{\mu\nu}^{(\lambda,1,1,0)})^*\Big).&
\end{eqnarray}
It seems that if one requires a full dS-covariant quantization, has to give up the positivity requirement of the inner product. Note that ($L\geq0$)
\begin{eqnarray}
&\langle h_{\mu\nu}^{(\lambda Llm)},h_{\mu\nu}^{(\lambda Llm)}\rangle=1,&\nn\\
&\langle (h_{\mu\nu}^{(\lambda Llm)})^*,(h_{\mu\nu}^{(\lambda Llm)})^*\rangle=-1.&
\end{eqnarray}

Following this path, applying the action of the dS group on the set of solution $h_{\mu\nu}^{(\lambda Llm)}$ ($L\geq0$) frequently, one can simply see that the smallest, complete, nondegenerate, and invariant inner product space for the gravitons field would be a Krein space
\begin{equation}\label{Krein space}{\cal{H}}= {\cal{H}}_+ + {\cal{H}}_-, \end{equation}
in which ${\cal{H}}_+$ is the Hilbert space constructed over the modes (\ref{0-mode}) and (\ref{first}),
\begin{eqnarray}\label{set1}
{\cal{H}}_+  = \hspace{6.7cm}\nn\\
\{\sum_{\lambda Llm,\; L\geq 0} c_{Llm} h_{\mu\nu}^{(\lambda Llm)}; \;\; \sum_{\lambda Llm,\; L\geq0} |c_{Llm}|^2<\infty\},\;\;
\end{eqnarray}
and ${\cal{H}}_-$ is an anti-Hilbert space (a negative definite inner product space). In other words, neither ${\cal{H}}_+$ nor ${\cal{H}}_-$ carry a representation of the dS group, so that, there is no covariant decomposition ${\cal{H}}_+ + {\cal{H}}_-$. However, the key point here is that the $SO(4)$-covariant decomposition exists. Note that, only the four generators of $M_{\alpha\beta}$ contracting to the spacetime translations, Eqs. (\ref{M01}) to (\ref{M04}), are responsible for de Sitter symmetry breaking. The other six generators associated with the compact $SO(4)$ subgroup, contracting to the Lorentz subalgebra, Eqs. (\ref{M12}) to (\ref{M43}), preserve de Sitter invariance and allow a $SO(4)$-covariant construction. Indeed, the set constituted of (\ref{first}) and (\ref{0-mode}), i.e. (\ref{set1}), is $SO(4)$-invariant, and by utilizing this set of modes, the $SO(4)$-covariant quantum field is quite available.

\section{The Quantum Field}
As already mentioned, the fully de Sitter-covariant gravitons field is expected to be an operator-valued distributions on $M_H$ acting on $\cal{H}$ (\ref{Krein space}). Let us recall that for any space $\cal{H}$  one defines the corresponding Fock space $\underline{\cal{H}}$  by
$$ \underline{\cal{H}} = \bigoplus_{n\geq0} S_n(\cal{H}), $$
where $S_n(\cal{H})$ is the $n$-th symmetrical tensor product of $\cal{H}$. When $\cal{H}$ is realized as a space $L^2({\mathbb{R}}^d,d\mu)$, one can realize $S_n(\cal{H})$ as the space of square integrable symmetric functions of $n$ variables on ${{\mathbb{R}}^d}$. The one-dimensional space $S_0(\cal{H})$ is written $|0\rangle$ and called the vacuum state. As is well known, the creators $a^{\dagger\lambda}_{Llm}$ and annihilators $a^\lambda_{Llm}$ create and annihilate, respectively, the mode $h_{\mu\nu}^{(\lambda Llm)}$. They can be realized on the Fock space in the following way
\begin{widetext}
\begin{eqnarray}\label{ann}
\Big(a(h)q\Big)(X_1,...,X_{n-1})=\sqrt{n} \frac{i}{H^2}\int_{\rho=0}\Big[{h}^*(\rho,\Omega)\cdot\cdot\partial_\rho{q}\Big((\rho,\Omega),X_1,...,X_{n-1}\Big)\hspace{4cm}\nn\\
- 2\frac{a-1}{a}((\partial_\rho x)\cdot{{h}}^*(\rho,\Omega))\cdot(\partial\cdot{q})\Big((\rho,\Omega),X_1,...,X_{n-1})\Big) - (1^* \leftrightharpoons 2)\Big]d\Omega,
\end{eqnarray}
for any square-integrable $n$-symmetric function $q$. The creator is defined as usual by
\begin{eqnarray}\label{crea}
\Big(a^\dagger(h)q\Big)(X_1,...,X_{n+1})=\frac{1}{\sqrt{n+1}} \sum_{i=1}^{n+1} {h}(X_i)\cdot\cdot{q}(X_1,...,\widetilde{X}_i,...,X_{n+1}),
\end{eqnarray}
where $\widetilde{X}_i$ means that this term is omitted. One can easily see that
\begin{eqnarray} \label{causality}
[a(h),a^\dagger(q)]=\langle h,q\rangle,
\end{eqnarray}
This gives, of course, the usual commutation relations when applied to the modes $h_{\mu\nu}^{(\lambda Llm)}$. One can also verify that
\begin{eqnarray}\label{gives}
\underline{U}_g a^\dagger(h) \underline{U}_g^* = a^\dagger(U_gh),\;\; \mbox{and} \;\; \underline{U}_g a(h) \underline{U}_g^* = a(U_gh).
\end{eqnarray}

We are now ready to define the (unsmeared) quantum field $\underline{h}_{\mu\nu}(X)$ on $\underline{\cal{H}}$ by (Note that, to respect the standard notation, once again, we introduce the quantum field by $\underline{h}_{\mu\nu}(X)$, which is obviously different from the previous one introduced in Sec. III.)
\begin{eqnarray}\label{qf}
\underline{h}_{\mu\nu}(X) = \sum_{\lambda Llm} a_{Llm}^\lambda h_{\mu\nu}^{(\lambda Llm)} - \sum_{\lambda Llm} b_{Llm}^\lambda (h_{\mu\nu}^{(\lambda Llm)})^* + \sum_{\lambda Llm}(a_{Llm}^\lambda)^\dagger (h_{\mu\nu}^{(\lambda Llm)})^* - \sum_{\lambda Llm} (b_{Llm}^\lambda)^\dagger h_{\mu\nu}^{(\lambda Llm)}, \end{eqnarray}
in which $a_{Llm}^\lambda$ and $b_{Llm}^\lambda$ are, respectively, the annihilators of the modes $h_{\mu\nu}^{(\lambda Llm)}$ and $(h_{\mu\nu}^{(\lambda Llm)})^*$. The nonvanishing commutation relations between these operators are
\begin{eqnarray}\label{ccr}
[a_{Llm}^\lambda , (a_{Llm}^{\lambda})^\dagger ] = 1,\;\;\;\; [b_{Llm}^\lambda , (b_{Llm}^{\lambda})^\dagger ] = -1.
\end{eqnarray}
It is worth mentioning that the minus sign follows from the formulas $ [a(h),a^\dagger(q)]=\langle h,q\rangle$ and the fact that $\langle h^*,h^*\rangle =-1$. Note also that this field is clearly real as the sum of an operator and its conjugate,
\begin{eqnarray}\label{newqf}
\underline{h}_{\mu\nu}(X) = \underline{h}^{(+)}_{\mu\nu}(X) + \underline{h}^{(-)}_{\mu\nu}(X),
\end{eqnarray}
where
\begin{eqnarray}\label{newqfpositive}
\underline{h}^{(+)}_{\mu\nu}(X) = \sum_{\lambda Llm} a_{Llm}^\lambda h_{\mu\nu}^{(\lambda Llm)} + \sum_{\lambda Llm}(a_{Llm}^\lambda)^\dagger (h_{\mu\nu}^{(\lambda Llm)})^* ,
\end{eqnarray}
and
\begin{eqnarray}\label{newqfnegative}
\underline{h}^{(-)}_{\mu\nu}(X) = - \sum_{\lambda Llm} b_{Llm}^\lambda (h_{\mu\nu}^{(\lambda Llm)})^*  - \sum_{\lambda Llm} (b_{Llm}^\lambda)^\dagger h_{\mu\nu}^{(\lambda Llm)}.
\end{eqnarray}
\end{widetext}

We claim that this field is covariant and causal.\\
In order to prove these claims, we proceed as in Eq. (\ref{pf}) to introduce the smeared field, which is easier to work with. We consider the distribution $p$ taking values in $\cal{H}$, so that, for any function $f$, $p(f)$ is the unique element of $\cal{H}$,
\begin{eqnarray}
\langle p(f) , q \rangle = (f,q), \;\; \; \forall q\in \cal{H}.
\end{eqnarray}
The existence of $p(f)$ is subject to a technical requirement on $\cal{H}$: the continuity for each $f$ of the map $q\mapsto (f,q)$, $\cal{H}\mapsto\mathbb{C}$. $p(f)=\int p^{\mu\nu}(X) f_{\mu\nu}(X) d\mu(X)$, therefore, leads to $\langle p(X),q\rangle=q(X)$. One can easily see that the kernel $\widetilde{G}$ of $p$ is given by
$$\langle p_{\mu\nu}(X),p_{\mu'\nu'}(X') \rangle = -i\widetilde{G}_{\mu\nu\mu'\nu'}(X,X'),$$
and the field can be written in a coordinate-free definition as follows
\begin{eqnarray}\label{pfnew}
\underline{h}_{\mu\nu}(X) = a(p_{\mu\nu}(X)) + a^\dagger(p_{\mu\nu}(X)).
\end{eqnarray}
The covariance of $\underline{h}_{\mu\nu}(X)$ now can be easily checked respecting (\ref{gives}) and the covariance of $p$.

The causality of this field now follows immediately from this definition and from the formula (\ref{causality}),
\begin{eqnarray}
[\underline{h}_{\mu\nu}(X),\underline{h}_{\mu'\nu'}(X')]&=&2\langle p_{\mu\nu}(X),p_{\mu'\nu'}(X')\rangle \nn\\
&=&-2i \widetilde{{G}}_{\mu\nu\mu'\nu'}(X,X').
\end{eqnarray}
The field is causal because $\widetilde{G}_{\mu\nu\mu'\nu'}(X,X')$ vanishes when $X$ and $X'$ are spacelike separated.

At the end, considering the above construction, the (Krein-Gupta-Bleuler) Fock vacuum is characterized by
$$ a^\lambda_{Llm}|0\rangle=b^\lambda_{Llm}|0\rangle=0, \;\;\;\;\; \forall L\geq0 \;\mbox{and}\; \lambda=1,...,10. $$
It is trivially invariant under the action of de Sitter group $SO_0(1,4)$.

Now, let us make explicit our result. Thus far, utilizing a robust group theoretical machinery, we have obtained a fully dS-covariant and causal quantization of the gravitons field on de Sitter background. The construction is, therefore, free of any infrared divergence. Our calculations clearly reveal that the only way to preserve the full dS covariance of the theory is including illegitimate negative norm states. In other word, there is no natural vacuum state (the Euclidean state) for free gravitons in de Sitter space that shares the background symmetries. To go round this difficulty, it seems that a restrictive version of covariance, in which Fock states include the Euclidian vacuum state as an invariant vacuum (not under the full dS group), should be considered. Insisting on the Euclidean vacuum stems from the fact that the existence of an invariant Euclidean vacuum as the natural dS vacuum state is essential to the notion of a de Sitter temperature \cite{Figari265} and the associated entropy \cite{Gibbons2738}. Therefore, we either have to redefine vacua invariant under a subgroup of the de Sitter group only (spontaneous symmetry breaking), or choose to restrict the field to a subset of the de Sitter spacetime, or consider invariance under the Lie algebra of the de Sitter group rather than under the full group action.

Respecting the above reasoning, in section III, at the same time with the main stream of our calculations in order to construct a fully dS-invariant set of modes, we have also obtained the set ${\cal{H}}_+$ (see (\ref{set1})) which is thoroughly invariant under a maximal subgroup of the dS group, namely $SO(4)$. Utilizing this family simply leads to the Euclidean vacuum in the standard terminology; considering this set of modes, the corresponding $SO(4)$-covariant quantum field, characterized by $\underline{h}^{(+)}_{\mu\nu}(X)$ (see (\ref{newqfpositive})), can be constructed over the Hilbertian Fock space, while the $SO(4)$-invariant Fock vacuum $|\widetilde{0}\rangle$ is given by
$$ a^\lambda_{Llm}|\widetilde{0}\rangle=0,  \;\;\;\;\; \forall L\geq0 \;\mbox{and}\; \lambda=1,...,10. $$

Before ending our discussions in this section, let us make an additional remark. Comment on the argument given by Woodard \emph{et al} in \cite{Woodard2009} about the gauge fixing procedure. They believe that the unjustified use of average gauge fixing is behind the mentioned dispute about the free gravitons in dS background. More accurately, they have reasoned that certain gauge fixing functionals cannot be added to the action on backgrounds such as de Sitter in which a linearization instability is present. In this regard, we must declare that there is no contradiction between our mathematical point of view in adding dS-invariant gauge fixing terms to the Lagrangian (\ref{L-expanded}) and the Woodard statement. In our investigations, indeed, by adding dS-invariant gauge fixing terms to the Lagrangian along with the use of the concrete structure of the dS group theory, we have presented a full dS-covariant quantization of the gravitons field. Again, the formalism is free of any infrared divergence, and obviously no linearization instability is present. Therefore, with respect to the Woodard viewpoint, there is no mathematical obstacle in our gauge fixing procedure. Of course, the price to pay in building the covariant quantum field, which is the appearance of un-physical negative norm states, forces us to give up dS invariance and so forth.

\section{Discussion}
Here, we should comment on the apparent conflict of our result with the argument given by the mathematical physics community maintaining that there is no physical breaking of de Sitter invariance. In this regard, let us focus on work by Higuchi \cite{higuchi}, which is a fundamental paper for this claim. The idea is that the de Sitter dynamical gravitons might be physically dS-invariant, although no manifestly dS-invariant propagator for them can be found \cite{Marolf&Morrison,Faizal&Higuchi}. This approach necessitates making sense of Bunch-Davies vacuum (as the unique possibility for a dS-invariant vacuum state for dynamical gravitons), and obviously, means somehow preventing the infrared divergence. In this paper, after a complete gauge fixing procedure, it is shown that the physical graviton modes can be chosen as $h_{\mu\nu}^{(\lambda Ll m)}$ with $L\geq l \geq 2$ and $\lambda = \pm$ (which is different from our notation and corresponds to the helicity). Under a de Sitter boost, these modes transform into other modes as well, but these other modes are of the form $\nabla_\mu \Xi_\nu + \nabla_\nu \Xi_\mu$. This means that, by defining the equivalence relation \begin{eqnarray}\label{transformation}
h'_{\mu\nu}\sim h_{\mu\nu} + \nabla_\mu \Xi_\nu + \nabla_\nu \Xi_\mu,
\end{eqnarray}
and by regarding $h_{\mu\nu}^{(\lambda Llm)}$ as the representative elements of the equivalence classes, the modes $h_{\mu\nu}^{(\lambda L l m)}$, $L \geq l \geq 2$, $-l \leq m \leq l$, form the unitary representation $\Pi_{2,2}^{+}\oplus \Pi_{2,2}^{-}$. Note that, the de Sitter group is represented on the space of solutions through satisfying the transverse-traceless-synchronous conditions $h_{0\mu} = 0$, $h^\mu_{\;\;\mu} = 0$ and $\nabla^\mu h_{\mu\nu} = 0$ in linearized gravity. Because the change from $h'_{\mu\nu}$ to $h_{\mu\nu}$ was a gauge transformation, Higuchi et al concluded that no observable quantity is affected. Because Bunch-Davies vacuum now exists, they conclude that the graviton vacuum is dS-invariant.

Regarding the above statement some crucial point should be clarified. On one hand, it must be underlined that the above argument by Higuchi suffers from fundamental defects coming from the gauge transformation. The gauge transformation (along with the transverse-traceless-synchronous conditions) alters both the propagator equation and the canonical commutation relations \cite{Miao0,Miao00}. The difficulty is that it seems to change things we have already measured. For instance, in Minkowski space QED, utilizing the same type of transformation seems to eliminate the infrared divergences of the exclusive scattering amplitudes which indeed have measurable consequences and cannot be changed. One can get a detailed discussion of these statements in \cite{Miao0,Miao00}.

On the other hand, if one respects the canonical commutation relations, we should declare that any canonical quantization procedure in which one only considers the invariant physical graviton modes does yield a non-covariant quantization of linearized gravity, and therefore, obtaining infrared finite graviton two-point function in this context is not surprising. We recall from the previous sections that the quantization of gauge-invariant theories, as is well known, usually requires quantization \`{a} \emph{la} Gupta-Bleuler which is based on three invariant spaces of solutions $V_g\subset V\subset V_a$ \cite{BinegarGUPTA,Gazeau1847}; the physical states space, the quotient space $V/V_g$ of states up to a gauge transformation for which the dS group acts through the unitary representation $\Pi^+_{2,2}\oplus\Pi^-_{2,2}$, is invariant but not invariantly complemented in $V_a$ \cite{PejhanI}. Indeed, an indecomposable group representation structure appears unavoidable, where the physical states belong to a subspace (characterized by the divergencelessness condition of the field operator) $V$ of solutions but where the field operator must be defined on a larger gauge dependent space $V_a$ (which contains negative norm states).\footnote{Again, this is deeply analogous to the case of the electromagnetic field in Minkowski space for which the only way to preserve (manifest) covariance and gauge invariance in canonical quantization is to use Gupta-Bleuler method \cite{JPA}.} Indeed, in gauge quantum field theories, it has been proven that the use of an indefinite metric is an unavoidable feature if one insists on the full covariance as well as the causality (locality) for the theory \cite{Strocchi}.

Frankly speaking, the correct procedure in all cases is to allow free gravitons to resolve their infrared problem by breaking de Sitter invariance.

\section*{Acknowledgements}
This work was partially supported by the JSPS KAKENHI Grant Number JP 25800136 and the research-funds presented by Fukushima University (K.B.).

\begin{appendix}
\setcounter{equation}{0}
\section{Mathematical relations underlying Eq. (\ref{mini2})}
In this appendix, the solution to the field equation (\ref{minimally}) is given. In this regard, it is convenient to rewrite (\ref{minimally}) in a more general form as follows
\begin{eqnarray}\label{A1}
Q_0\phi=\kappa\phi,\;\;\; \kappa=\langle Q_0\rangle={(\frac{m_H}{H})}^2 + 12\xi,
\end{eqnarray}
in which $m_H$ and $\xi$, respectively, refer to a ``mass" and a positive gravitational coupling with the dS background. Note that, the Laplace-Beltrami operator, $\Box$, and the scalar part of the Casimir operator, $Q_0$, are proportional,
$$\Box = - H^{2}Q_0.$$
Therefore, the equation (\ref{A1}) simply leads to the wave equation for scalar fields propagating on dS spacetime,
\begin{eqnarray}\label{A2}
[\Box + (m_H^2 + 12H^2\xi)]\phi=0.
\end{eqnarray}

The Laplace-Beltrami operator on dS spacetime is
\begin{eqnarray}\label{Laplace}
\Box= H^2\cos^4\rho\frac{\partial}{\partial\rho}(\cos^{-2}\rho\frac{\partial}{\partial\rho})-H^2 \cos^2\rho\Delta_3,
\end{eqnarray}
where
\begin{eqnarray}\label{Laplace'}
\Delta_3&=&\frac{\partial^2}{\partial\alpha^2} + 2\cot\alpha\frac{\partial}{\partial\alpha} + \frac{1}{\sin^2\alpha}\frac{\partial^2}{\partial\theta^2}\nn\\
&+& \cot\theta\frac{1}{\sin^2\alpha}\frac{\partial}{\partial\theta} + \frac{1}{\sin^2\alpha\sin^2\theta}\frac{\partial^2}{\partial\varphi^2},
\end{eqnarray}
is the Laplace operator on the hyperbolic $S^3$.

Considering $\phi(X)=U(\rho)V(\Omega),\; \Omega\in S^3$ (separation of variable), Eq. (\ref{A1}) or equivalently (\ref{A2}) can be divided into two parts,
\begin{eqnarray}\label{Laplace-V}
(\Delta_3 + C)V(\Omega)=0,
\end{eqnarray}
\begin{eqnarray}\label{Laplace-U}
(\cos^4\rho\frac{d}{d\rho}\cos^{-2}\rho \frac{d}{d\rho} + C \cos^2\rho + \kappa)U(\rho)=0.
\end{eqnarray}
Concentrating on the angular part, for $C=L(L+2)$ $L\in\mathbb{N}$, we obtain $V={\cal{Y}}_{Llm}$,
\begin{eqnarray}\label{hyperspherical}
{\cal{Y}}_{Llm}(\Omega)&=& \Big(\frac{(L+1)(2l+1)(L-1)!}{2\pi^2(L+l+1)!} \Big)^{\frac{1}{2}}\nn\\
&\times &2^l l! (\sin\alpha)^lC_{L-1}^{l+1}(\cos\alpha)Y_{lm}(\theta,\varphi),
\end{eqnarray}
in which $0\leq l\leq L,\;\; 0\leq |m|\leq l$, the $C^{\lambda}_{n}$ are Gegenbauer polynomials \cite{Talman} and
\begin{eqnarray}\label{hyperspherical}
Y_{lm}(\theta,\varphi)=(-1)^m \Big( \frac{(l-m)!}{(l+m)!}\Big)^{\frac{1}{2}}P_l^m(\cos\theta)e^{im\varphi},
\end{eqnarray}
$P_l^m$ are the corresponding Legendre functions. ${\cal{Y}}_{Llm}$'s satisfy the orthogonality conditions
$$\int_{S^3} {\cal{Y}}_{Llm}(\Omega) {\cal{Y}}_{L'l'm'}(\Omega) d\Omega=\delta_{LL'} \delta_{ll'} \delta_{mm'},$$
$d\Omega =\sin^2\alpha \sin\theta d\alpha d\theta d\phi$ is the $O(4)$-invariant measure on $S^3$.

On the other hand, following \cite{Kirsten}, the radial part (\ref{Laplace-U}) would be
\begin{eqnarray}\label{radial}
U_{\lambda L}(\rho)=A_L(\cos\rho)^{\frac{3}{2}} [ P^\lambda_{L+\frac{1}{2}}(\sin\rho) - \frac{2i}{\pi}Q^\lambda_{L+\frac{1}{2}}(\sin\rho)],\;\;\;
\end{eqnarray}
with
\begin{eqnarray}\label{lambda}
\lambda&=&\sqrt{\frac{9}{4} - \kappa} \;\;\mbox{when}\;\frac{9}{4}\geq\kappa\geq 0, \nn\\
\lambda&=&i\sqrt{\kappa - \frac{9}{4}} \;\;\mbox{when}\;\frac{9}{4}\leq\kappa.
\end{eqnarray}
$P^{\lambda}_{n}$ and $Q^{\lambda}_{n}$ stand for Legendre functions on the cut, and
\begin{eqnarray}
A_L=H \frac{\sqrt{\pi}}{2} \Big( \frac{\Gamma(L-\lambda+\frac{3}{2})}{\Gamma(L+\lambda+\frac{3}{2})} \Big)^{\frac{1}{2}}.
\end{eqnarray}
The complete set of modes for the field equation (\ref{A1}) or equivalently (\ref{A2}) then would be
\begin{eqnarray}\label{general}
\phi^\lambda_{Llm}=U_{\lambda L}(\rho) {\cal{Y}}_{Llm}(\Omega), \;\; X=(\rho,\Omega)\in M_H
\end{eqnarray}
There exists an exception, the massless minimally coupled scalar field $\kappa=0$, for which the above construction breaks down (for $L=0$). However, giving up $L=0$, the above formulas are still valid for $\kappa=0$ and one can simply obtain the field solution to (\ref{minimally}) as (\ref{mini2}).
\end{appendix}

\end{document}